\documentclass{article}

\usepackage[final]{neurips_2020}

\usepackage[utf8]{inputenc} % allow utf-8 input
\usepackage[T1]{fontenc}    % use 8-bit T1 fonts
\usepackage{hyperref}       % hyperlinks
\usepackage{url}            % simple URL typesetting
\usepackage{booktabs}       % professional-quality tables
\usepackage{amsfonts}       % blackboard math symbols
\usepackage{nicefrac}       % compact symbols for 1/2, etc.
\usepackage{microtype}      % microtypography

\usepackage{amsmath}
\usepackage{graphicx}
\usepackage{array}
\newcolumntype{C}[1]{>{\centering\arraybackslash}m{#1}}
\usepackage{subfigure}

\title{ESMFL: Efficient and Secure  Models for Federated Learning}

\author{
    Sheng Lin\\
    Northeastern University\\
    lin.sheng@northeastern.edu\\
    \And
    Chenghong Wang\\
    Duke University\\
    cw374@duke.edu\\
    \And
    Hongjia Li\\
    Northeastern University\\
    hongjia.li@northeastern.edu\\
    \And
    Jieren Deng\\
    University of Connecticut\\
    li.hongjia@northeastern.edu\\
    \And
    Yanzhi Wang\\
    Northeastern University\\
    yanz.wang@northeastern.edu\\
    \And
    Caiwen Ding\\
    University of Connecticut\\
    caiwen.ding@uconn.edu\\
}

\begin{document}

\maketitle

\begin{abstract}
Nowadays, Deep Neural Networks are widely applied to various domains. However, massive data collection required for deep neural network reveals the potential privacy issues and also consumes large mounts of communication bandwidth. To address these problems, we propose a privacy-preserving method for the federated learning distributed system, operated on Intel Software Guard Extensions, a set of instructions that increase the security of application code and data. Meanwhile, the encrypted models make the transmission overhead larger. Hence, we reduce the commutation cost by sparsification and it can achieve reasonable accuracy with different model architectures.

% Experimental results under our privacy-preserving framework show that, for LeNet-5, we obtain 98.78\% accuracy on IID data and 97.60\% accuracy on Non-IID data with 34.85\% communication saving, and 1.8$\times$ total elapsed time acceleration. For MobileNetV2, we obtain 85.40\% accuracy on IID data and 81.66\% accuracy on Non-IID data with 15.85\% communication saving, and 1.2$\times$ total elapsed time acceleration.
\end{abstract}

\section{Introduction}
Large-scale deep neural networks (DNNs) introduce intensive computation and high memory storage, bringing challenges on current edge devices (clients, e.g, mobile phones) with limited resources~\cite{he2016deep, krizhevsky2012imagenet}.
% consist of at least millions to hundreds of millions of parameters and training data~\cite{he2016deep, krizhevsky2012imagenet, deng2009imagenet},
% 
Thus, large-scale DNN models as well as training data usually are stored on centralized cloud server clusters in data center~\cite{wang2010privacy, bonawitz2019towards}. However, data privacy and security has been increasingly concerned in cloud servers, where the sensitive data are either owned by vendors or customers~\cite{wang2010privacy}. 
% It increases the risk of users' privacy exposure to the party that they don't want to share. 
Federated learning (FL) has been developed for DNN training without acquiring raw data from the users~\cite{bonawitz2019towards,hard2018federated}. It is a distributed machine learning approach which enables training
on a large corpus of decentralized data on edge devices (large in number) and only collects the local model or gradient for global synchronization on a central server~\cite{mcmahan2017federated}. Through the local training, FL enhances data privacy. 

% It enables edge devices to collaboratively train a shared global model under the coordination of a central server.
However, there are still some challenges when it comes to FL. (i) Edge devices typically have a limited communication bandwidth and computation resources
% mobile devices usually have much lower computational power 
compared to the server. Therefore, training large-scale DNNs will consume a large amount of communication time and resources~\cite{mcmahan2016communication,jiang2019model}. (ii) Traditional FL method can not guarantee data privacy. Recent research~\cite{DBLP:journals/corr/abs-1906-08935,mcmahan2017learning} shows that the publicly shared gradients in training process can reveal private information to either a third-party, or a central server.

% To make the communication from the edge devices to the central server efficient, 

Model compression techniques such as weight pruning~\cite{jiang2019model} and weight quantization~\cite{konevcny2016federated} have been introduced into FL, to reduce the number of parameters or  bit-representation communicated at each round. To enhance the privacy of FL, current works typically use the classical cryptographic protocols such as differential privacy, i.e., randomly perturbing the intermediate results (adding noise) at each iteration~\cite{agarwal2018cpsgd,geyer2017differentially}. The noise will introduce perturbation on the FL model, leading to accuracy degradation in overall accuracy. To make it worse, adding noise makes a sparse model to a dense model, and it is not compatible with weight pruning techniques.

% schemes such as sparsification, subsampling, and quantization can significantly reduce the size of messages
% communicated at each round.
% Current works that aim to improve the privacy of FL typically build upon previous classical cryptographic protocols such as differential privacy~\cite{agarwal2018cpsgd,geyer2017differentially}.

On the other hand, a standard implementation of the federated learning system requires that multiple clients train full models with their own data, then the server aggregates the model parameters in each round. However, most high-speed Internet connections, including cable, digital subscriber line (DSL) and fiber, are asymmetric. Due to higher downstream demand, high speed Internet providers have designed their systems to provide much better speed for downloading than uploading. Therefore, for transmitting large models, the bottleneck of FL communication cost is mainly restricted by the uploading. 
% Therefore, if we want to reduce the overall communication cost of a complete round, we need reduce both communication cost from uploading and downloading.

% \textcolor{red}{There are some aforementioned studies~\cite{} that specifically focus on optimizing the performance of FL algorithms, however, those algorithms only reduce the communication cost without considering the potential privacy issues. As ~\cite{liu2019enhancing} mentioned, even when user data remains local, there are still a lot of threats to user privacy.}

To overcome these challenges, it is important to enhance the performance of FL under the secured model training. We investigate privacy-preserving methods which can preserve the overall accuracy, reduce the communication cost and accelerate overall elapsed time. In this paper, we propose a framework, ESMFL, operated on Software Guard Extensions (SGX). ESMFL integrates weight pruning with federated learning that can reduce the communication cost efficiently while preserving data privacy. 
Our main contributions are as follows:
\begin{itemize}
\item We propose a systematic method to set the regularization penalty coefficient that enables parameter regularization and accelerates sparse model training by incorporating ADMM-based pruning algorithm.

\item We enhance the security of the gradient and model in federated learning process by integrating SGX, a set of instructions that increases the security of application code and data , giving them more protection from disclosure or modification.

\item ESMFL can achieve a higher accuracy with same training rounds compared to prior arts. It continuously reduce the communication cost and reconfigure its architecture into more cost-efficient.

\end{itemize}

We evaluate our framework on two datasets, MNIST and CIFAR-10 with analysis on model sparsity, communication cost and elapsed time on each modules.
For LeNet-5 on MNIST, we obtain 98.78\% (98.6\%) accuracy on IID data with 9.99$\times$ (87$\times$) model compression rate and 97.60\% (94.53\%) accuracy on Non-IID data with same model compression rate as IID data. When the model compression rate is 9.99$\times$ (87$\times$), communication is saved by 34.85\% (48.78\%) and total elapsed time is reduced by 1.8$\times$ (2.4$\times$). For MobileNetV2 on CIFAR-10, we obtain 85.40\% (81.23\%) accuracy on IID data with 4.95$\times$ (8.7$\times$) model compression rate and 81.66\% (80.44\%) accuracy on Non-IID data with same model compression rate as IID data. When the model compression rate is 4.95$\times$ (8.7$\times$), communication cost is saved by 15.85\% (31.3\%) and total elapsed time is reduced by 1.2$\times$ (1.8$\times$).

\section{ESMFL Framework}
% \textcolor{red}{In this section, we describe the modules in ESMFL’s workflow (Section 3.1),workflow (Section 3.2), and trust assumptions (Section 3.3). (delete this)}
% \subsection{FL Framework with Masked Pruning}\label{section:flmp}

% % A canonical round of federated training includes three phases: (1) The server sends the global model to a subset of the edge devices; (2) Each client device trains the model using private local data; (3) The server averages the local updates. 

% To The efficient communication and update restricts the shared weights to have a sparse structure. As the baseline of our paper, we restrict the local updates to be a sparse matrix, and we prune the weights based on the magnitude, i.e, setting a mask to map the lowest percent portion of weights to zeros. It is important to stress that we train the updates of this structure in each round and each client independently. We set thresholds in each layer in neural networks, and achieve a certain overall compression rate according to the neural network architecture.

% Figure Overview.\\

\begin{figure}[t]
    \centering
    \includegraphics[width=0.6 \textwidth]{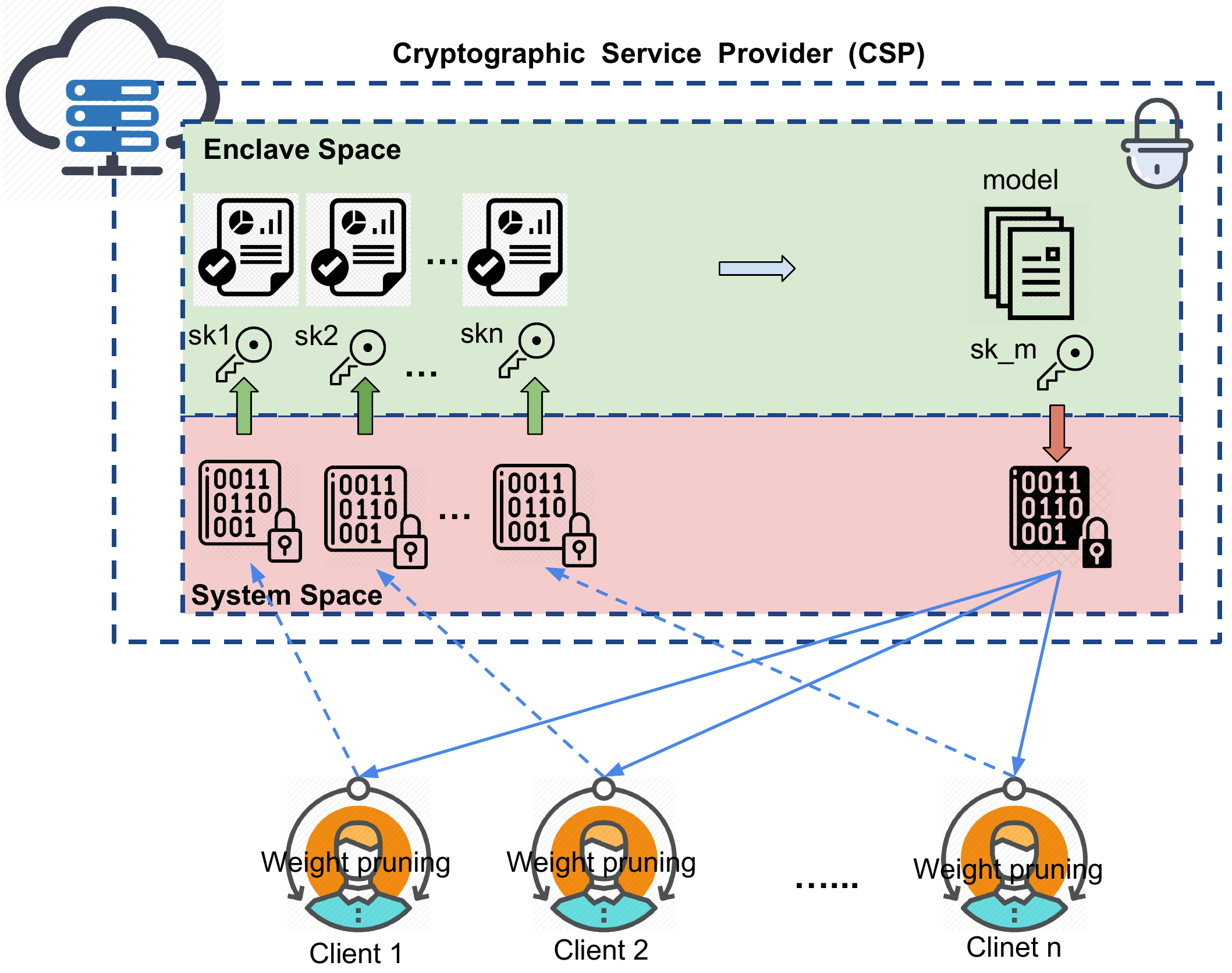}
    %\vspace{-1mm}
    \caption{The proposed ESMFL framework.}
    \label{fig:framework}
    % \vspace{-3mm}
\end{figure}

\subsection{Modules in ESMFL}

ESMFL employs one untrusted Cryptographic Service Provider (CSP) that runs the SGX and a group of clients who conduct local training and submit local updates.  The \textit{CSP} initializes and manages the cryptographic primitives, and aggregate model updates each round within Enclave space. The following contents describe the detailed modules of each participant.
\subsubsection{Cryptographic Service Provider (CSP)}

The \textit{Key Manager} module locates within enclave space. \textit{Key Manager} initializes the symmetric encryption key for each client via remote attestation~\cite{anati2013innovative} and stores them to decrypt encrypted updates in each training round.

The \textit{CSP} is the only entity capable of aggregate model submitted by each client. The \textit{Model Aggregation} module is tasked with decryption and handling all updates within the trusted space (SGX Enclave), and publish new models according to the aggregated updates.

\subsubsection{Client}

 The \textit{Local Trainer} trains the model based on the private data owned by the client. During the local training, the model compression algorithms are applied to \textit{Local Trainer} to obtain a sparse model after the training process is done.
 
% \textcolor{blue}{Please add descriptions here}\\
 The \textit{Data Encryption} module stores the update encryption key
%  $k_i$ 
 of client 
%  $C_i$  
 which is negotiated with the \textit{CSP} via remote attestation. Each client 
%  $C_i$ 
 encrypt their local updates in each round using  the encryption key 
%  $k_i$ 
 and sends the encrypted update to the \textit{CSP} via a secure channel. 

\subsection{ESMFL Workflow}

Figure~\ref{fig:framework} shows the overall workflow of our ESMFL system. At the very outset, the \textit{CSP} initializes a trusted execution space (i.e., Intel SGX Enclave) and waits for attestation requests. Clients who wish to contribute to the federated training process (denoted as $C_i$) then make a remote attestation request to CSP. If a valid attestation is provided by \textit{CSP}, the client negotiates a symmetric encryption key $sk_i$. After all clients complete the remote attestation process, \textit{CSP} initialize a weight matrix with random entries and broadcast all clients. Clients download the global initial model then apply several epochs of training using local data and obtain a local model (we use $\Delta_i$ to denote the update for client $C_i$) for current training epoch. The clients then encrypt their update, $\Delta_i$, in the binary format with key $sk_i$ and send the encrypted update (denoted as $\hat{\Delta_i}$) to \textit{CSP}, which load them into the attested enclave. Next, \textit{CSP} decrypts $\hat{\Delta_i}$ using corresponding $sk_i$ stored in the \textit{Key Manager}. \textit{CSP} keep collecting updates from all clients, decrypt and aggregate them to a single one, which uses federated averaging method to obtain the updated model. \textit{CSP} then publish the updated model to clients and clients repeat the local training process. The training terminates when certain condition hits. According to such workflow, the local update $\Delta_i$ of each client $C_i$ in each epoch is only observable to the client itself and the attested enclave on \textit{CSP}. Therefore for any computational bounded adversary, there is no possibility to investigate the $\Delta_i$ of $C_i$.

\subsection{Weight Pruning in Clients}\label{section:admmpr}
As the baseline for comparison, we use sparse matrices to do the local updates, and call this method federated average masked pruning. We prune the local weights before sending to CSP based on the magnitude. It is important to stress that we train the updates of this structure in each round and each client independently. We set thresholds in each layer in neural networks, and achieve a certain overall compression rate according to the neural network architecture.

% , i.e, setting a mask to map the lowest percent portion of weights to zeros.

Posing a direct mask to the client updates is lack of the regularization of structure, which cannot be efficiently used in each round. Thus, it is hard to achieve extremely high compression rate with reasonable accuracy and fast coverage speed by this way. Considering the performance of weight pruning, we develop a FL framework with the state-of-the-art Alternating Direction Method of Multipliers (ADMM) based pruning algorithm~\cite{zhang2018systematic}. The sparse model FL training process can be divided into two phases: warm-up training and weight pruning. 

The objective of the warm-up training is to train the model without compression for initial several rounds. Since in the early stages of training, the parameters in neural network are changing rapidly. According to our preliminary experiments, it is better to pruning based on a well-trained model than pruning from scratch and it can achieve a better converge performance than pruning from scratch. 

In the second phase, the objective of the local weight pruning is to minimize the loss function while satisfying the constraints of weight sparsity. In the local client, we define the weight pruning problem in clients as:  
\vspace{-0.20em}
\begin{equation}\label{eqn:goal}
\centering
\begin{aligned}
{\text{minimize\ }} \bf{f}_{L} \big( \{{{W}}_{i}\}_{i=1}^N, \{{{b}}_{i} \}_{i=1}^N \big),\;  \\
\text{subject \; to\; }
{\bf{{W}}}_{i}\in {\mathcal \bf{S} }_{i},\  i = 1, \ldots, N,
\end{aligned}
\end{equation}
where ${W}_{i}$ and ${b}_{i}$ denotes the sets of weights and biases of the $i$-th (CONV or FC) layer in an $N$-layer DNN, respectively. The set ${\mathcal{S}}_{i}=\big\{{{W}_{i}\big|\text{card}({{W}}_{i})\le n_i\big\}}$ denotes the constraint for weight pruning, and `card' refers to cardinality. It meets the goal that the number of non-zero elements in ${{W}}_{i}$ is limited by $n_i$ in layer $i$. 

In the local weight pruning phase, we add the ADMM-based regularization~\cite{zhang2018systematic} on all original DNN models. The regularization is operated by introducing auxiliary variables ${{Z}}_{i}$'s, and dual variables ${{U}}_{i}$'s. It proceed by step $s = 0, 1, 2\ldots$, iteratively until convergence as the following subproblems:
\vspace{-0.20em}
\begin{equation}
\small
    \bf{W}_{i}^{s+1} := \underset{ {\bf{W}}_{i}}{\text{arg min}}\quad \bf{F}_{L}(\{\bf{W}_i\}, \{\bf{Z}_i^s\}, \{\bf{U}_i^s\}),
\label{itera1}
\end{equation}
\vspace{-0.50em}
\begin{equation}
\small
    \bf{Z}_{i}^{s+1} := \underset{ {\bf{Z}}_{i}}{\text{arg min}}\ \bf{F}_{L}(\{\bf{W}_i^{s+1}\}, \{\bf{Z}_i\}, \{\bf{U}_i^s\}),
\label{itera2}
\end{equation}
\vspace{-0.50em}
\begin{equation}
\small
    \bf{U}_{i}^{s+1} := \bf{U}_{i}^{s}+\bf{W}_{i}^{s+1}-\bf{Z}_{i}^{s+1}.
\label{itera3}
\end{equation}

% The detail process  is shown in Algorithm~\ref{algorithmTraining}.

In each iteration, while keeping on minimizing network regularized loss, we also reduce the error of Euclidean projection from  ${{W}}_{i}^{k+1}+{{U}}_{i}^{k}$ onto the set ${\mathcal{S}}_{i}$. Because under the constraint that $\alpha_i$ is the desired number of weights after pruning in the $i$-th layer, the Euclidean projection can keep $\alpha_i$ elements in ${{W}}_{i}^{k+1}+{{U}}_{i}^{k}$ with the largest magnitudes and set the remaining weights to zeros. Then the dual variables ${{U}}_{i}$ is updated as following: ${U}^{k+1}_{i} = {{U}}^{k}_{i} + {{W}}^{k+1}_{i} - {{Z}}^{k+1}_{i}$. After repeating several steps, we obtain the trained intermediate ${{W}}_{i}$. Finally, we perform the Euclidean projection to map weights to configured sparsity ratio that at most $\alpha_i$ weights in each layer are non-zero. 

% \begin{algorithm}[htp]
% \footnotesize
% \textbf{Input:} ;\\
% \textbf{Initialization:} ; \\
 
% \textbf{Return:} .
% \caption{ADMM based pruning}
% \label{alg:pruning}
% \end{algorithm}

% \begin{algorithm}[t]
% \caption{Local Client Weight pruning based on ADMM}
% \footnotesize
% \label{algorithmTraining}
% \SetAlgoLined
%     Synchronize models from the server\;
%     Initialize hyperparameters in a local client\;
%     \For {Current\_Epoch $<$ MAX\_Local\_Epoch\_for\_One\_Round\_FL}
%     {
%         Solve Subproblem (Eqn. (\ref{itera1}))\;
%         \If{Current\_Epoch $\%$ ADMM\_Interval\_Epoch == 0}{
%         Solve Subproblem (Eqn. (\ref{itera2}))\;
%         Dual variable update according to Eqn. (\ref{itera3})\;}
%     }
%     Map to the configured mask\;
%     Upload pruned model to the server\; 
    
% \end{algorithm}{}

% And then in the retraining phase, the zero weights are gradient masked and non-zero weights are retrained using training sets to further improve accuracy.

\section{Experiments}

\subsection{Experimental Setup}
We implement the baseline and our proposed framework by PyTorch~\cite{paszke2017automatic} and simulate multiple clients and a CSP with different FL settings on a server with a 3.1GHz Intel Xeon Scalable Processors (8 virtual CPU with 32 GB memory) and a NVIDIA 2080 GPU (8 GB memory).

\textbf{Datasets and Models}: 
In our experiments, we test LeNet5~\cite{lecun2015lenet} for MNIST dataset, MobileNetV2~\cite{sandler2018mobilenetv2} for CIFAR-10. To study the federated optimization, we also need to specify how the data is distributed over the clients. We use the similar dataset setting as~\cite{mcmahan2016communication} described. We partition the MNIST dataset in two different ways over clients, namely IID and non-IID. In the MNIST with IID data over clients, the whole data is shuffled, and then divided into 100 clients with 600 examples per client balancedly. In the MNIST with non-IID data over clients, we sort the whole data by label index, then divide it into 200 shards of size 300, and assign each of 100 clients 2 shards. Similarly, we partition the CIFAR-10 dataset into IID and non-IID over clients. In the CIFAR-10 with IID data over clients, the whole data is shuffled, and then divided into 100 clients each receiving 500 examples. These partitions are balanced. In the CIFAR-10 with non-IID data over 100 clients, we sort the whole data by label index, then divide it into 500 shards of size 100, and assign each client with 5 shards.

\begin{table*}[b]
\centering
\setlength{\abovecaptionskip}{0pt}%    
\setlength{\belowcaptionskip}{4pt}%
\caption {Accuracy and model specifications} \label{tab:accspar}
\resizebox{\textwidth}{!}{%
\renewcommand{\arraystretch}{1.2}% for the vertical padding
\setlength\tabcolsep{1.5pt}
\begin{tabular}{|c|c|c|c|c|c|c|c|}
% \begin{tabular}{|C{1.5cm}|C{1.5cm}|C{1cm}|C{1cm}|C{1cm}|C{3.5cm}|C{2.5cm}|C{2.5cm}|}
\hline
\textbf{Network} & \textbf{Dataset} & \textbf{Params} & \textbf{Accuacy} &\textbf{Data} &\textbf{Percentage Non-Zero Weights} &\textbf{Compression Rate}& \textbf{CSP Pruning Round} \\ \hline
LeNet-5           &  MNIST                &     430K            &     98.78\% &  IID          &  10.01\%  & 9.99 & 50                                \\ \hline
LeNet-5           &   MNIST               &     430K            &    98.26\%   & IID          &  1.15\%  & 87.0 &50                                \\ \hline
LeNet-5           &  MNIST                &     430K            &     97.60\% &  non-IID          &  10.01\%  & 9.99 &  50                                \\ \hline
LeNet-5           &   MNIST               &     430K            &    94.53\%   & non-IID          &  1.15\%  & 87.0 &50                                \\ \hline
MobileNetV2      &    CIFAR-10                &   2.28M              &    85.40\% & IID             &  20.2\%   &   4.95 & 100                                  \\ \hline
MobileNetV2      &  CIFAR-10                &  2.28M              &     81.23\% & IID          &   11.49\%   &  8.70 & 100                                 \\\hline
MobileNetV2      &    CIFAR-10                &   2.28M              &    81.66\% & non-IID             &  20.2\%   &   4.95 & 100                                  \\ \hline
MobileNetV2      &  CIFAR-10                &  2.28M              &     80.44\% & non-IID          &   11.49\%   &  8.70 & 100                                 \\\hline
\end{tabular}
}
\end{table*}

\begin{table*}[t]
\centering
\setlength{\abovecaptionskip}{0pt}%    
\setlength{\belowcaptionskip}{4pt}%
\caption {Data communication volume and percentage of communication cost} \label{tab:comcost}
\resizebox{\textwidth}{!}{%
\renewcommand{\arraystretch}{1.2}% for the vertical padding
\setlength\tabcolsep{1.5pt}
\begin{tabular}{|c|c|c|c|c|}

% \begin{tabular}{|C{1.8cm}|C{1.5cm}|C{4cm}|C{4cm}|C{4.5cm}|}
\hline
\textbf{Network} & \textbf{Dataset} & \textbf{Data Communication Volume (Non-Sparse)} & \textbf{Data Communication Volume (Sparse)} & \textbf{Percentage of Communication Cost Saving} \\ \hline
Lenet-5           & MNIST            &       1720MB                                          &     1120.58MB                                     &          34.85\%                                 \\ \hline
Lenet-5           & MNIST            &       1720MB                                           &  881.02MB                                         &        48.78\%                                   \\ \hline
MobileNetV2      & CIFAR10          &  18.24GB                                               &     15.38GB                                        &        15.68\%                                   \\ \hline
MobileNetV2      & CIFAR10          &    18.24GB                                             &      12.53GB                                       &     31.3\%     \\ \hline                                
\end{tabular}
}
\end{table*}

\begin{figure}[t]
    \centering
    \includegraphics[width=1 \textwidth]{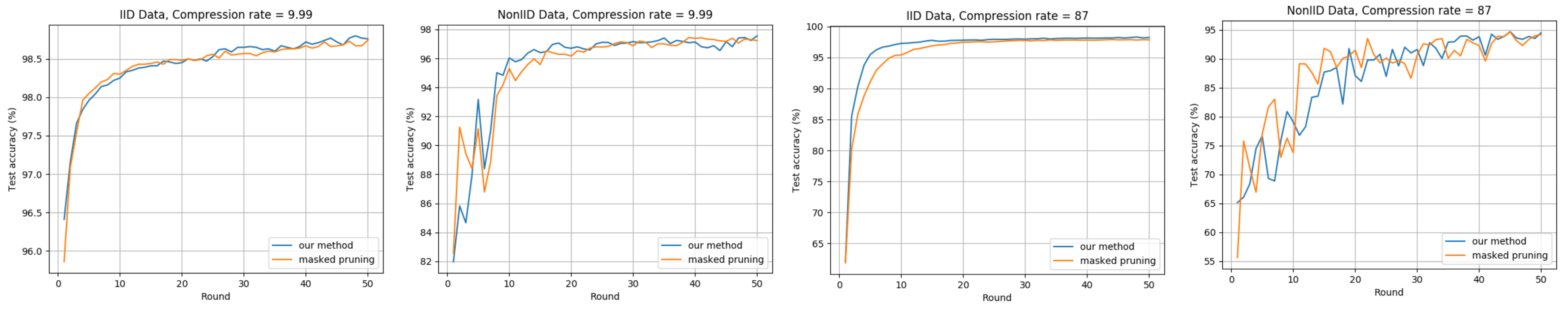}
    %\vspace{-1mm}
    \caption{Sparse model federated training process evaluation for LeNet-5 on MNIST with (a) IID and CR=9.99, (b) Non-IID and CR=9.99, (c) IID and CR=87  and (d) Non-IID and CR=87.}
    \label{fig:mnist_all}
    % \vspace{-3mm}
\end{figure}

\begin{figure}[t]
    \centering
    \includegraphics[width=1 \textwidth]{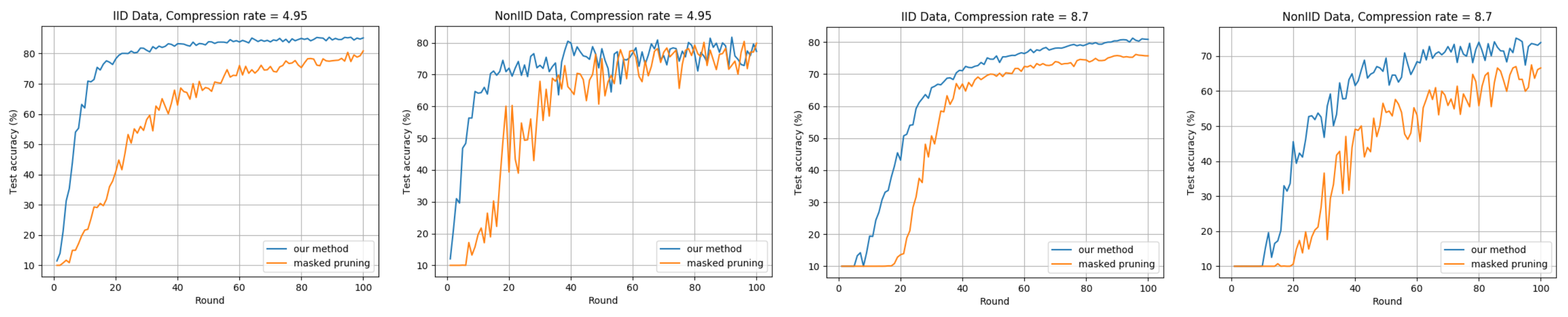}
    %\vspace{-1mm}
    \caption{Sparse model federated training process evaluation for MobileNetV2 on CIFAR-10 with (a) IID and CR=4.95, (b) Non-IID and CR=4.95, (c) IID and CR=8.7 and (d) Non-IID and CR=8.7.}
    \label{fig:cifar_all}
    % \vspace{-3mm}
\end{figure}

% \begin{figure}[t]
%     \centering
%     \includegraphics[width=1 \textwidth]{figures/fed_cifar10_mobilenetv2_v.png}
%     %\vspace{-1mm}
%     \caption{Sparse model federated training process evaluation for MobileNetV2 on CIFAR-10 with (a) IID and CR=4.95, (b) IID and CR=8.7, (c) Non-IID and CR=9.99 and (d) Non-IID and CR=8.7.}
%     \label{fig:cifar_all}
%     % \vspace{-3mm}
% \end{figure}

% \begin{figure}[t]
%     \centering
%     \includegraphics[width=1 \textwidth]{figures/fed_mnist_lenet5_v.png}
%     %\vspace{-1mm}
%     \caption{Sparse model federated training process evaluation for LeNet-5 on MNIST with (a) IID and CR=9.99, (b) IID and CR=87, (c) Non-IID and CR=9.99 and (d) Non-IID and CR=87.}
%     \label{fig:mnist_all}
%     % \vspace{-3mm}
% \end{figure}

\subsection{Model Sparsity}
We start by investigating the effects of our proposed weight pruning methods on MNIST. In the first experiment, we divide the MNIST data in IID partition. The performance analysis on the effects of different compression rates on the convergence speed and test accuracy of ESMFL and Federated Averaging masked pruning method can be found in Figure~\ref{fig:mnist_all}. As mentioned in Section~\ref{section:admmpr}, for both our method and the baseline method, we first warm-up train the model in non-sparse way to achieve higher accuracy, then we start pruning process. In Figure~\ref{fig:mnist_all}(a) and Figure~\ref{fig:mnist_all}(c), we observe when the compression rate is low, our pruning method and masked pruning perform similarly in MNIST dataset. Our method achieves 98.78\% accuracy on 89.99\% sparsity in 50 FL pruning rounds. However, as we increase the pruning ratio in weight update, our method achieve faster convergence speed and even achieve better accuracy. Our method achieves 98.26\% accuracy on 98.85\% sparsity in 50 FL pruning rounds, and masked pruning achieves 97.92\% in the same sparsity. In Figure~\ref{fig:mnist_all}(b) and Figure~\ref{fig:mnist_all}(d),  for the non-IID partition, our method achieves 97.60\% accuracy on 89.99\% sparsity and 94.53\% accuracy on 98.85\% sparsity compared with masked pruning with best accuracy, 97.36\% and 94.52\% on corresponding sparsities.

\begin{table*}[]
\centering
\setlength{\abovecaptionskip}{0pt}%    
\setlength{\belowcaptionskip}{4pt}%
\caption {Breakdown elapsed time evaluation for each  modules} \label{tab:performance}
\resizebox{\textwidth}{!}{%
\renewcommand{\arraystretch}{1.2}% for the vertical padding
\setlength\tabcolsep{1.5pt}
\begin{tabular}{|c|c|c|c|c|c|c|}
% \begin{tabular*}{\textwidth}{@{\extracolsep{\fill}}|c|c|c|c|c|c|c|}
% \begin{tabular}{|C{6.2cm}|C{1.45cm}|C{1.45cm}|C{1.45cm}|C{1.45cm}|C{1.45cm}|C{1.45cm}|}
\hline
\textbf{Task}   &\multicolumn{3}{c|}{\textbf{MNIST on LetNet-5}} & \multicolumn{3}{c|}{\textbf{CIFAR-10 on MobileNetV2} } \\ \hline
\textbf{Data Size (MB)}                            & 172.00                 & 52.12                  & 4.20                   & 912.00                       & 621.76                       & 340.56                       \\ \hline
\textbf{Client-Server Attestation Time (s)}            & 2.2883                 & 2.3410                 & 2.3027                 & 2.2040                       & 2.2374                       & 2.2790                       \\ \hline
\textbf{Data Provisioning / Round (s)}             & 1.1785                 & 0.7810                 & 0.6217                 & 4.2521                       & 3.3079                       & 1.9331                       \\ \hline
\textbf{Network Transmission Time / Round (s)}     & 8.4556                 & 4.4233                 & 3.0346                 & 36.2894                      & 27.2418                      & 15.0892                      \\ \hline
\textbf{Ecall Time / Round (s)}                     & 2.5146                 & 0.3980                 & 0.1593                 & 15.7002                      & 11.7166                      & 5.6256                       \\ \hline
\textbf{Ocall Time / Round (s)}                            & 2.4728                 & 0.5731                 & 0.0638                 & 14.5418                      & 10.4928                      & 5.3021                       \\ \hline
\textbf{Local Training Time / Round (s)}                            & 0.8457                 & 1.2366                & 1.2476                 & 21.5948                      & 21.7753                      & 22.1206                       \\ \hline
\textbf{Global Model Aggregation Time / Round (s)}                            & 0.008                 & 0.008                 & 0.008                 & 0.3351                       & 0.3272                      & 0.3461                      \\ \hline
\textbf{Total Time / Round (s)}                            & 17.7635                 & 9.7610                 & 7.4377                 & 94.9174                      & 77.0990                      & 52.6757                       \\ \hline

\end{tabular}
}
\end{table*}

For the IID partition of the CIFAR-10 data, we choose the MobileNetV2 as our model for comparisons. The MobileNetV2 is a neural network architecture specially designed for mobile devices, therefore it is a good candidate model for FL framework. The FL training process of MobileNetV2 is same as MNIST. As the result shown in Figure~\ref{fig:cifar_all}(a), our method achieves 85.40\% accuracy on 79.8\% sparsity in 100 FL pruning rounds compared with masked pruning, which results in 80.88\% accuracy with the same sparsity. As the result shown in Figure~\ref{fig:cifar_all}(c), our method achieves 81.23\% accuracy on 88.51\% sparsity in 100 FL pruning rounds and masked pruning achieves 76.12\% accuracy in the same sparsity. As the result shown in Figure~\ref{fig:cifar_all}(b) and Figure~\ref{fig:cifar_all}(d) for the non-IID partition, our method achieves 81.66\% accuracy on 79.80\% sparsity and 75.15\% accuracy on 88.51\% sparsity compared with masked pruning with best accuracy, 80.44\% and 67.38\% on corresponding sparsities. Our sparse model test accuracy result and model specifications are summarized in Table~\ref{tab:accspar}. It is obvious that our method performs better than the baseline in CIFAR-10, since the architecture of MobileNetV2 is more complex and deeper than LetNet-5.

\subsection{Communication Cost and Performance} 

The data communication volume is the total transmission data volume during the federated training process. Here, we calculate the data communication volume only for weight pruning phase. we encode the model in the compressed sparse row (CSR) format. The communication cost for total 50 rounds on MNIST and total 100 rounds on CIFAR-10 are summarized in Table~\ref{tab:comcost}. From the table, we observe it can achieve at least 34.85\% (15.68\%) communication cost reduction compared with standard non-compressed way on MNIST (CIFAR-10). If we further compress the update in each federated training round, we can achieve 48.88\% ( 31.30\%) communication cost reduction on MNIST (CIFAR-10). 

To get a better evaluation of the performance of our ESMFL we increase the participating client number to 100 in each round federated training. The breakdown elapsed time evaluations for the  performance of each module are summarized in Table~\ref{tab:performance}. We record the elapsed time in the whole flow in our ESMFL framework as five parts. \textit{Client-Server Attestation Time:} The total time for remote attestation and key dissemination between enclave and the client. \textit{Data Provisioning Time:} The total time for clients encrypted the updated model obtained from local training. \textit{Network Transmission Time:} The time for network exchange between CSP and Clients (time is per round and per client based). \textit{Ecall Time:} The total time that leverages ECALLs (i.e. load encrypted local updates into enclave, etc.) \textit{OCALL Time:} The total time that leverages OCALLs (i.e. after update the global model, export from enclave and publish.)

Using ESMFL, we can achieve performance improvement 1.8$\times$ (2.4$\times$) with 10$\times$ (87$\times$) compression rate on MNIST and  1.2$\times$ (1.8$\times$) with 4.95$\times$ (8.7$\times$) compression rate on CIFAR-10 on a federated training round. For small model like LeNet-5 on MNIST, the bottleneck is the local train time, and for larger model like MobileNetV2, reducing the Ecall Time and Ocall Time become more critical to overall performance.

\section{Conclusion}
In this paper, we propose a privacy-preserving method for the federated learning distributed system, which can reduce the commutation cost and achieve a reasonable accuracy with extreme sparse model. 
Our federated learning system ensures the data privacy without adding noise. We reduce the communication cost compared to unencrypted federated learning system while maintaining the overall accuracy by applying weight pruning. Compared to prior weight pruning methods in federated learning system, our method can achieve much higher pruning ratio at the same accuracy level.

\section{Acknowledgement} 
This work is funded by the National Science Foundation Awards CNS-1739748 and CNS-1704662.

% \begin{ack}
% Use unnumbered first level headings for the acknowledgments. All acknowledgments
% More information about this disclosure can be found at: \url{https://neurips.cc/Conferences/2020/PaperInformation/FundingDisclosure}.

% Do {\bf not} include this section in the anonymized submission, only in the final paper. You can use the \texttt{ack} environment provided in the style file to autmoatically hide this section in the anonymized submission.
% \end{ack}

% \bibliographystyle{unsrtnat}

% \bibliographystyle{abbrv}
% \bibliography{reference}
\bibliography{neurips_2020}
% \section*{References}

% References follow the acknowledgments. Use unnumbered first-level heading for
% the references. Any choice of citation style is acceptable as long as you are
% consistent. It is permissible to reduce the font size to \verb+small+ (9 point)
% when listing the references.
% {\bf Note that the Reference section does not count towards the eight pages of content that are allowed.}
% \medskip

% \small

% [1] Alexander, J.A.\ \& Mozer, M.C.\ (1995) Template-based algorithms for
% connectionist rule extraction. In G.\ Tesauro, D.S.\ Touretzky and T.K.\ Leen
% (eds.), {\it Advances in Neural Information Processing Systems 7},
% pp.\ 609--616. Cambridge, MA: MIT Press.

% [2] Bower, J.M.\ \& Beeman, D.\ (1995) {\it The Book of GENESIS: Exploring
%   Realistic Neural Models with the GEneral NEural SImulation System.}  New York:
% TELOS/Springer--Verlag.

% [3] Hasselmo, M.E., Schnell, E.\ \& Barkai, E.\ (1995) Dynamics of learning and
% recall at excitatory recurrent synapses and cholinergic modulation in rat
% hippocampal region CA3. {\it Journal of Neuroscience} {\bf 15}(7):5249-5262.

\end{document}